\documentclass[journal]{IEEEtran}

\usepackage{graphicx}
\usepackage{subfigure}
\usepackage{amsmath}
\usepackage{amssymb}
\usepackage{textcomp}
\usepackage{cite}
\usepackage{url}
\usepackage{color}
\usepackage[normalem]{ulem}

\newcommand{\pf}[1]{\texttt{#1}} 

\newcommand{\units}[1]{\; \text{#1}}
\newcommand{\e}[1]{\times 10^{#1}}

\begin{document}

\title{Online Training of an Opto-Electronic Reservoir Computer Applied to Real-Time Channel Equalisation}

\author{Piotr~Antonik, Fran\c{c}ois~Duport, Michiel~Hermans, Anteo~Smerieri, Marc~Haelterman, and Serge~Massar%
\thanks{Piotr Antonik, Michiel Hermans and Serge Massar are with the Laboratoire d'Information Quantique, Universit\'{e} Libre de Bruxelles, 50 Avenue F. D. Roosevelt, CP 224, 1050 Brussels, Belgium.}%
\thanks{Fran\c{c}ois Duport, Anteo Smerieri and Marc Haelterman are with the Service OPERA-Photonique, Universit\'{e} Libre de Bruxelles, 50 Avenue F. D. Roosevelt, CP 194/5, 1050 Brussels, Belgium.}%
\thanks{Correspondence and requests for materials should be addressed to Piotr Antonik (pantonik@ulb.ac.be).}%
\thanks{This is the author's version of an article that has been published in IEEE TNNLS. Changes were made to this version by the publisher prior to publication. The final version of record is available at http://dx.doi.org/10.1109/TNNLS.2016.2598655.}%
\thanks{\copyright \, 2016 IEEE. Personal use of this material is permitted. Permission from IEEE must be obtained for all other uses, in any current or future media, including reprinting/republishing this material for advertising or promotional purposes, creating new collective works, for resale or redistribution to servers or lists, or reuse of any copyrighted component of this work in other works.}%
}

\maketitle

\begin{abstract}
  Reservoir Computing is a bio-inspired computing paradigm for processing time dependent signals. The performance of its analogue implementation are comparable to other state of the art algorithms for tasks such as speech recognition or chaotic time series prediction, but these are often constrained by the offline training methods commonly employed. Here we investigated the online learning approach by training an opto-electronic reservoir computer using a simple gradient descent algorithm, programmed on an FPGA chip. Our system was applied to wireless communications, a quickly growing domain with an increasing demand for fast analogue devices to equalise the nonlinear distorted channels. We report error rates up to two orders of magnitude lower than previous implementations on this task. We show that our system is particularly well-suited for realistic channel equalisation by testing it on a drifting and a switching channels and obtaining good performances.
\end{abstract}


\IEEEpeerreviewmaketitle

\section{Introduction}

\IEEEPARstart{R}{eservoir} Computing (RC) is a set of methods for designing and training artificial recurrent neural networks \cite{jaeger2001echo, maass2002real} that brings a drastic simplification of the system design. A typical reservoir is a randomly connected fixed network with arbitrary coupling coefficients between the input signal and the nodes. These parameters remain fixed and only readout weights are optimised. This greatly simplifies the training process - that is, computing the coefficients of the readout layer - which often reduces to solving a system of linear equations. Despite these simplifications, the RC approach can yield performances equal, or even better than other machine learning algorithms \cite{lukovsevivcius2009survey, hammer2009recent, lukovsevivcius2012reservoir, lukovsevivcius2012practical}. The RC algorithm has been applied to speech and phoneme recognition, equalling other approaches \cite{verstraeten2006reservoir, jaeger2007optimization, triefenbach2010phoneme}, and won an international competition on financial time series prediction \cite{NFC}.

Optical computing has been investigated for decades as photons propagate faster than electrons, without generating heat or magnetic interference, and thus promise higher bandwidth than conventional computers \cite{arsenault2012optical}.
The possibility of optical implementation of reservoir computing was studied using numerical simulations in \cite{vandoorne2008toward}. A major breakthrough occurred by the end 2011 – beginning 2012 – when experimental implementations of reservoir computers with performance comparable to state of the art digital implementations were reported. In quick succession appeared an electronic implementation \cite{appeltant2011information}, and then three opto-electronic implementations \cite{paquot2012optoelectronic, larger2012photonic, martinenghi2012photonic}. Since then all-optical reservoir computers have been reported using as nonlinearity the saturable gain of a semiconductor optical amplifier \cite{duport2012all}, a semiconductor laser with delayed feedback \cite{brunner2012parallel}, the saturation of absorption \cite{dejonckheere2014all}, integrated on an optical chip \cite{vandoorne2014experimental}, and based on a coherently driven passive optical cavity \cite{vinckier2015high}.

The performance of a reservoir computer greatly relies on the training technique used to compute the readout weights. Offline learning methods, used up to now in experimental implementations \cite{vandoorne2008toward, appeltant2011information, paquot2012optoelectronic, larger2012photonic, martinenghi2012photonic, duport2012all, brunner2012parallel, dejonckheere2014all, vandoorne2014experimental}, provide good results, but become detrimental for real-time applications, as they require large amounts of data to be transferred from the experiment to the post-processing computer. This operation may take longer than the time it takes the reservoir to  process the input sequence \cite{paquot2012optoelectronic, duport2012all, dejonckheere2014all}. Moreover, offline training is only suited for time-independent tasks, which is not always the case in real-life applications. The alternative (and more biologically plausible) approach is to progressively adjust the readout weights using various online learning algorithms such as gradient descent, recursive least squares or reward-modulated Hebbian learning \cite{bottou1998online}. Such procedures require minimal data storage and have the advantage of being able to deal with a variable task: should any parameters of the task be altered during the training phase, the reservoir computer would still be able to produce good results by properly adjusting the readout weights. 

In the present work we apply this online learning approach to an opto-electronic reservoir computer and show that our implementation is well suited for real-time data processing. The system is based on the opto-electronic reservoir, introduced in \cite{paquot2012optoelectronic, larger2012photonic}, coupled to an FPGA chip, that implements input and output layers. It generates the input sequence in real time, collects the reservoir states and computes optimal readout weights using a simple gradient descent algorithm. Real-time generation of reservoir inputs allows the system to be trained and tested on an arbitrary long input sequence, and the replacement of the personal computer by a dedicated FPGA chip significantly reduces the experimental runtime. We apply our system to a specific real-world task: the equalisation of nonlinear communication channel.

Wireless communications is by far the fastest growing segment of the communications industry. The increasing demand for higher bandwidths requires pushing the signal amplifiers close to the saturation point which, in turn, adds significant nonlinear distortions into the channel. These have to be compensated by a digital equaliser on the receiver side \cite{benedetto1999principles}. The main bottleneck lies in the Analog-to-Digital Converters (ADCs) that have to follow the high bandwidth of the channel with sufficient resolution to sample correctly the distorted signal \cite{singh2009multi}. Current manufacturing techniques allow producing fast ADCs with low resolution, or slow ones with high resolution, obtaining both being very costly.
This is where analog equalisers become interesting, as they could equalise the signal before the ADC and significantly reduce the required resolution of the converters, thus potentially cutting costs and power consumption \cite{sobel2009a, feng2010new, yong201160ghz}. Moreover, optical devices may outperform digital devices in terms of processing speed \cite{hassan2010analog, sobel2009a}. It can for instance be shown that reservoir computing implementations can reach comparable performance to other digital algorithms (namely, the Volterra filter \cite{malone2011practical}) for equalisation of a nonlinear satellite communication channel \cite{bauduin2015equalization}.

Our reservoir computer is used to equalise a simple wireless channel introduced in \cite{mathews1994adaptive}. This model is described by a simple set of equations (see section \ref{subsec:cheq}) and can be easily implemented on the FPGA chip. This task has also been extensively studied in the RC community, both numerically \cite{jaeger2004harnessing} and experimentally \cite{paquot2012optoelectronic, duport2012all, dejonckheere2014all, vinckier2015high}. Our system performs better than previously reported RC implementations on this task and we report error rates up to two orders of magnitude lower than previous results \cite{paquot2012optoelectronic, duport2012all, dejonckheere2014all, vinckier2015high}. Furthermore, we demonstrate the great advantage of online training, namely that it is suitable for solving non-stationary tasks, such as a variable wireless channel. This is particularly interesting for real-life applications, as physical communication channels vary depending on fluctuating environmental conditions. We show that even under such variable conditions, our system performs as well as in the stationary case.

In previous work we programmed the simple gradient descent algorithm on an FPGA chip to train a digital reservoir computer \cite{antonik2015fpga}, and we have reported preliminary results on an online-trained physical reservoir computer \cite{antonik2015online}. Compared to the latter work, the experimental setup has been improved, the FPGA design has been further optimised, and a new dedicated clock generation device is used. As a consequence the system is more stable, more efficient, and the reservoir size has been increased to 50 neurons (as in \cite{paquot2012optoelectronic, duport2012all, dejonckheere2014all, vinckier2015high}). We also report what is, to the best of our knowledge, the lowest error rates ever obtained with a physical reservoir computer on the channel equalisation task. Finally we present a much more in depth analysis of the time-dependent case.

The paper is structured as follows. Section \ref{sec:basic} introduces the basic principles of the reservoir computing, the channel equalisation task and the simple gradient descent algorithm. The experimental setup and the FPGA design are outlined in sections \ref{sec:expsetup} and \ref{sec:design}. Finally, the experimental results and the conclusion are presented in sections \ref{sec:results} and \ref{sec:ccl}.

\section{Basic principles}
\label{sec:basic}

\subsection{Reservoir Computing}
\label{subsec:rc}

\begin{figure}
  \centering
  \includegraphics[width=0.45\textwidth]{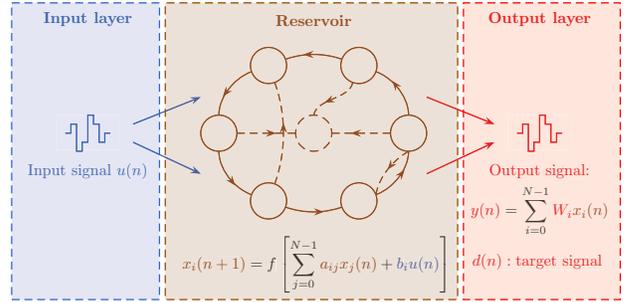}
  \caption{Schematic representation of a reservoir computer. Brown lines represent a general reservoir with random interconnections, solid lines highlight a reservoir with ring topology, used here. The time multiplexed input signal $u(n)$ is injected into a dynamical system, composed of a large number $N$ of internal variables $x_i(n)$. The dynamics of the system is defined by the nonlinear function $f$ and the coefficients $a_{ij}$ and $b_i$. The readout weights $w_i(n)$ are trained to obtain an output signal $y(n)$, given by their linear combination with the reservoir states $x_i(n)$, as close as possible to the target signal $d(n)$.}
  \label{fig:rc}
\end{figure}

A typical reservoir computer is depicted in figure \ref{fig:rc}. It contains a large number $N$ of internal variables $x_i(n)$ evolving in discrete time $n \in \mathbb{Z}$, as given by
\begin{equation}
  x_i(n+1) = f \left( \sum_{j=0}^{N-1} a_{ij} x_j(n) + b_i u(n) \right),
  \label{eq:rcevo}
\end{equation}
where $f$ is a nonlinear function, $u(n)$ is some external signal that is injected into the system, and $a_{ij}$ and $b_i$ are time-independent coefficients, drawn from some random distribution with zero mean, that determine the dynamics of the reservoir. The variances of these distributions are adjusted to obtain the best performances on the task considered.

The nonlinear function used here is $f = \sin (x)$, as in \cite{larger2012photonic, paquot2012optoelectronic}. To simplify the interconnection matrix $a_{ij}$, we exploit the ring topology, proposed in \cite{rodan2011minimum}, so that only the first neighbour nodes are connected. This architecture provides performances comparable to those obtained with complex interconnection matrices, as demonstrated numerically in \cite{lukovsevivcius2009survey} and experimentally in \cite{appeltant2011information,larger2012photonic,paquot2012optoelectronic,duport2012all,brunner2012parallel}. Under these circumstances we obtain
\begin{subequations}
  \begin{align}
    x_0(n+1) & = \sin \left(  \alpha x_{N-1}(n-1) + \beta M_0 u(n) + \phi \right),\label{eq:rcevo2_1} \\
    x_i(n+1) & = \sin \left(  \alpha x_{i-1}(n) + \beta M_i u(n) + \phi \right),\label{eq:rcevo2_2}
  \end{align}%
  \label{eq:rcevo2}%
\end{subequations}
with $i=1,\ldots,N-1$, $\alpha$ and $\beta$ parameters are used to adjust the feedback and the input signals, respectively, and $M_i$ is the input mask, drawn from a uniform distribution over the the interval $[-1, +1]$, as in \cite{rodan2011minimum, paquot2012optoelectronic, duport2012all}. A bias $\phi$ is used to shift the sine function from its symmetric point to compensate for the asymmetric channel output symbol distribution, as explained in section \ref{subsubsec:cheqconst}.

The reservoir computer produces an output signal $y(n)$, given by a linear combination of the states of its internal variables
\begin{equation}
  y(n) = \sum_{i=0}^{N-1} w_i x_i (n),
\end{equation}
where $w_i$ are the readout weights, trained either offline (using standard linear regression methods), or online, as described in section \ref{subsec:gd}, in order to minimise the square error between the output signal $y(n)$ and the target signal $d(n)$.

\subsection{Channel equalisation task}
\label{subsec:cheq}

The channel equalisation task \cite{mathews1994adaptive, jaeger2004harnessing, rodan2011minimum, boccato2011echo, boccato2012extended, rodan2012simple}, in addition to its practical interest, doesn't require the use of large reservoirs to obtain state-of-the-art results \cite{paquot2012optoelectronic, duport2012all, dejonckheere2014all, vinckier2015high}.

\subsubsection{Channel model}
\label{subsubsec:cheqconst}

The channel input signal $d(n)$ contains 2-bit symbols with values picked randomly from $\{-3, -1, 1, 3\}$. The channel is modelled by a linear system with memory of length 10 \cite{mathews1994adaptive}
\begin{align}
  \begin{split}\label{eq:qn}
    q(n) {}& = 0.08 d(n+2) - 0.12 d(n+1) + d(n) \\
           & + 0.18 d(n-1) - 0.1 d(n-2) + 0.091 d(n-3) \\
           & - 0.05 d(n-4) + 0.04 d(n-5) + 0.03 d(n-6) \\
           & + 0.01 d(n-7),
  \end{split}
\end{align}
followed by an instantaneous memoryless nonlinearity
\begin{equation}
  u(n) = q(n) + 0.036 q^2(n) - 0.011q^3(n) + \nu(n),
  \label{eq:chan}
\end{equation}
where $u(n)$ is the channel output signal and $\nu(n) = A \cdot r(n)$ is the added noise of amplitude $A$, where $r(n)$ is drawn from a uniform distribution over the interval $\left[-1, +1\right]$ (for ease of implementation on an FPGA chip). Noise amplitude values $A$ are chosen to produce the same signal-to-noise ratios as in \cite{paquot2012optoelectronic, duport2012all}, where Gaussian noise was used. The reservoir computer has to restore the clean signal $d(n)$ from the distorted noisy signal $u(n)$. The performance is measured in terms of wrongly reconstructed symbols, called the Symbol Error Rate (SER). The results are presented in section \ref{subsec:constch} and compared to a previous implementation based on the same opto-electronic setup.

Note that although the input signal $d(n)$ has a symmetric symbol distribution around $0$, the output signal $u(n)$ loses this property, with the symbols lying within the $\left[ -2.8, 4.5 \right]$ interval. The equaliser must take this shift into account and correct the symbol distribution properly.

\subsubsection{Influence of channel model parameters on equaliser performance}

\begin{figure}
  \centering
  \includegraphics[width=0.44\textwidth]{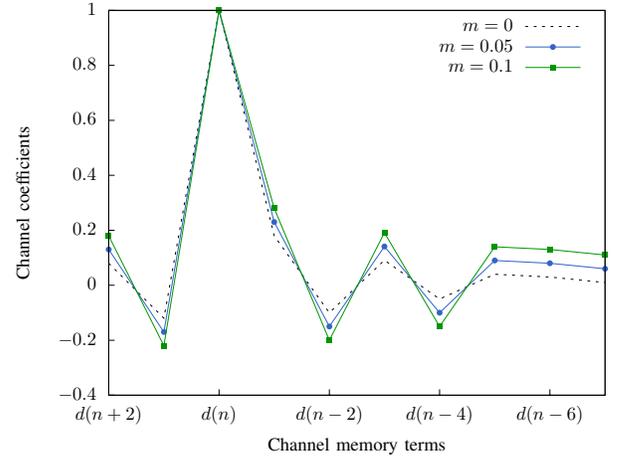}
  \caption{Various channel impulse responses, given by equation \eqref{eq:qngen}, for different values of $m$. Note that the $d(n)$ coefficient is kept fixed at $1$. Dotted curve shows the default shape defined by equation \eqref{eq:qn}.}
  \label{fig:mmshape}
\end{figure}

Equations \eqref{eq:qn} and \eqref{eq:chan} model a particular channel with certain amounts of symbol interference and nonlinear distortion, defined by the numerical values of the coefficients employed. To obtain a better understanding of this particular channel model, and to show which stages of input signal distortion are the most difficult to equalise, we introduce a more general channel model, given by
\begin{align}
  \begin{split}\label{eq:qngen}
    q(n) {}& = (0.08+m) d(n+2) - (0.12+m) d(n+1) \\
           & + d(n) + (0.18+m) d(n-1) \\
           & - (0.1+m) d(n-2) + (0.091+m) d(n-3) \\
           & - (0.05+m) d(n-4) + (0.04+m) d(n-5) \\
           & + (0.03+m) d(n-6) + (0.01+m) d(n-7),
  \end{split} \\
  \begin{split}\label{eq:changen}
    u(n) {}& = p_1 q(n) + p_2 q^2(n) + p_3 q^3(n),
  \end{split}
\end{align}
and we investigate the equalisation performance for different values of parameters $p_i$ and $m$. To preserve the general shape of the channel impulse response we keep the coefficient of $d(n)$ fixed at $1$ in equation \eqref{eq:qngen}. Figure \ref{fig:mmshape} shows the resulting impulse responses, given by equation \eqref{eq:qngen}, for several values of $m$. The results of these investigations are presented in the Appendix.

\subsubsection{Slowly drifting channel}
\label{subsec:cheqdrift}

The model given by equations \eqref{eq:qn} and \eqref{eq:chan} describes an idealistic stationary noisy wireless communication channel, that is, the channel remains the same during the transmission. However, in wireless communications, the environment has a great impact on the received signal. Given its highly variable nature, the properties of the channel may be subject to important changes in real time. 

To investigate this scenario, we performed a series of experiments with a ``drifting'' channel model, where parameters $p_i$ or $m_i$ were varying in real time during the signal transmission. These variations occurred at slow rates, much slower than the time required to train the reservoir computer. We studied two variation patterns: a monotonic increase (or decrease) and slow oscillations between two fixed values. Section \ref{subsec:driftchan} shows the results we obtained with our implementation.

\subsubsection{Switching channel}
\label{subsec:cheqswitch}

In addition to slowly drifting parameters, the channel properties may be subject to abrupt variations due to sudden changes of the environment. For better practical equalisation performance, it is crucial to be able to detect significant channel variations and adjust the RC readout weights in real time. We consider here the case of a ``switching'' channel, where the channel model switches instantaneously. The reservoir computer has to detect such changes and automatically trigger a new training phase, so that the readout weights get adapted for the equalisation of the new channel.

Specifically, instead of a constant channel, given by equations \eqref{eq:qn} and \eqref{eq:chan}, we introduce three channels differing in nonlinearity
\begin{subequations}
  \begin{align}
    u_1(n) & = 1.00 q(n) + 0.036 q^2(n) - 0.011q^3(n), \label{eq:ch1} \\
    u_2(n) & = 0.80 q(n) + 0.036 q^2(n) - 0.011q^3(n), \label{eq:ch2} \\
    u_3(n) & = 0.60 q(n) + 0.036 q^2(n) - 0.011q^3(n), \label{eq:ch3}
  \end{align}%
  \label{eq:varch}%
\end{subequations}
and switch regularly from one channel to another, keeping equation \eqref{eq:qn} unchanged. The results of this experiment are presented in section \ref{subsec:swch}.

\subsection{Gradient descent algorithm}
\label{subsec:gd}

This section describes the basic idea of the online training algorithm used here and introduces two modifications we investigated in our new implementation.

The gradient, or steepest, descent method is an algorithm for finding a local minimum of a function using its gradient \cite{arfken1985mathematical}. For the channel equalisation task considered here, the rule for updating the readout weights is given by \cite{bishop2006pattern}
\begin{equation}
  w_i (n+1) = w_i (n) + \lambda \left( d(n) - y(n) \right) x_i (n),
  \label{eq:wt}
\end{equation}
where $\lambda$ is the step size, used to control the learning rate. At high values of $\lambda$, the weights get close to the optimal values very quickly (in a few steps), but keep oscillating around these values. At low values, the weights converge slowly to the optimal values. In practice, we start with a high value $\lambda=\lambda_0$, and then gradually decrease it during the training phase until a minimum value $\lambda_{min}$ is reached, according to the equation
\begin{equation}
  \lambda(m+1) = \lambda_{min} + \gamma \left( \lambda(m) - \lambda_{min} \right),
  \label{eq:lambda_evo}
\end{equation}
with $\lambda(0) = \lambda_0$ and $m= \left\lfloor n/k \right \rfloor$, where $\gamma<1$ is the decay rate and $k$ is the update rate for the parameter $\lambda$.

The gradient descent algorithm suffers from a relatively slow convergence towards the global minimum, but its simplicity, with few simple computational steps, and flexibility, as the convergence rate and the resulting performance can be improved by tuning the parameters $\lambda$ and $\gamma$, make it a reasonable choice for a first implementation on a FPGA chip. Future investigations may focus on other online training algorithms, such as recursive least squares \cite{haykin2000adaptive} (a more computationally intensive method that converges faster) or unsupervised learning \cite{legenstein2010reward} (which doesn't require exact knowledge of the target output, but only an estimation of the reservoir performance).

\subsubsection{Full version}
\label{subsubsec:gdfull}

The step size parameter $\lambda$ is used to control the learning rate, and can also be employed to switch the training on or off. That is, setting $\lambda$ to zero stops the training process. This is how experiments on a stationary channel are performed: $\lambda$ is programmed to decay from $\lambda(0)$ to $0$ during a defined period, and then the reservoir computer performance is tested over a sequence of symbols, with constant readout weights. 

\subsubsection{Non-stationary version}
\label{subsubsec:gdnonstationary}

When equalising a drifting channel, the reservoir should be able to follow the variations and adjust the readout weights accordingly. This can be achieved by setting $\lambda_{min} > 0$ and thus letting the training process continue during the drift of the channel parameters. This procedure was used for experiments described in section \ref{subsec:driftchan}.

\subsubsection{Simplified version}
\label{subsubsec:simptrain}

As mentioned in the previous paragraph, the equalisation of a non-stationary channel requires keeping $\lambda_{min} > 0$. However, this worsens the equalisation performance, as the readout weights keep oscillating around the optimal values. This can be seen from equation \eqref{eq:wt}, that defines the update rule for the readout weights: at each time step $n$, a small correction $\Delta w_i = \lambda(n) (d(n) - y(n)) x_i(n) $ is added to every weight $w_i$. These corrections are gradually reduced by decreasing the learning rate $\lambda(n)$, so that the weights converge to their asymptotic values. In the case of a constant $\lambda$, the corrections $\Delta w_i$ are only damped by the error $d(n)-y(n)$, which stops decreasing at some point, leaving the $w_i$ oscillating around the optimal values.

To check the impact of a constant $\lambda$ on the equalisation performance we performed several experiments with a simplified version of the training algorithm by setting $\gamma=0$, and hence $\lambda(n) = \lambda_0$ for all $n$. Although this method will increase the error slightly, it has several advantages. With $\lambda$ constant, there is no need to search for an optimal decay rate $k$, which results in fewer experimental parameters to scan and thus shorter overall experiment runtime. Keeping $\lambda$ at a constant, non-zero value would also allow the equaliser to follow a drifting channel, as described in section \ref{subsec:cheqdrift}. The results obtained with this simplified version of the algorithm are shown in section \ref{subsec:ressimptrain}.

\section{Experimental setup}
\label{sec:expsetup}

\begin{figure}
  \centering
  \includegraphics[width=0.45\textwidth]{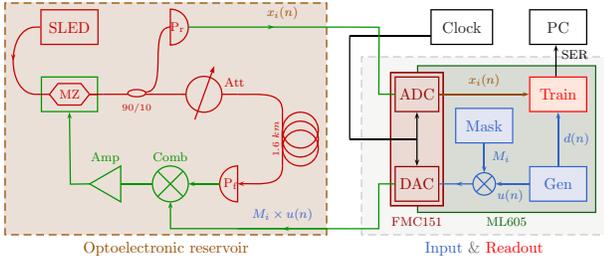}
  \caption{Schematic representation of the experimental setup. It contains an incoherent light source (SLED), a Mach-Zehnder intensity modulator (MZ), a $90/10$ beam splitter, an optical attenuator (Att), an approximately $1.6 \units{km}$ fibre spool, two photodiodes ($\text{P}_\text{r}$ and $\text{P}_\text{f}$), a resistive combiner (Comb) and an amplifier (Amp). The optical and electronic components are shown in red and green, respectively. The FPGA board implements both the input and output layers, generating the input symbols and training the readout weights. The computer controls the devices and records the results.}
  \label{fig:exp}
\end{figure}

Our experimental setup is depicted in figure \ref{fig:exp}. It contains three distinctive components: the optoelectronic reservoir, the FPGA board implementing the input and the readout layers and the computer used to setup the devices and record the results. The following sections present detailed overviews of these components, and section \ref{subsec:expparams} outlines the experimental parameters, tuned to obtain the best results.

\subsection{Optoelectronic reservoir}

The optoelectronic reservoir is based on the same scheme as in \cite{paquot2012optoelectronic, larger2012photonic}. These implementations use essentially the same hardware, but differ as to whether a low-pass filter is present in the cavity, and whether the input is desynchronised with respect to the cavity roundtrip. We use here the desychronised version of \cite{paquot2012optoelectronic}, without low-pass filter. The reservoir states are encoded into the intensity of incoherent light signal, produced by a superluminiscent diode (Thorlabs SLD1550P-A40). The Mach-Zehnder (MZ) intensity modulator (Photline MXAN-LN-10) implements the nonlinear function, its operating point is adjusted by applying a bias voltage, produced by a Hameg HMP4040 power supply.  A fraction (10\%) of the signal is extracted from the loop and sent to the readout photodiode and the resulting voltage signal is sent to the FPGA. The optical attenuator (JDS HA9) is used to set the feedback gain $\alpha$ of the system (see equations \eqref{eq:rcevo2}). The fibre spool consists of approximately $1.6 \units{km}$ single mode fibre, giving a round trip time of $7.94 \units{\textmu s}$. The resistive combiner sums the electrical feedback signal, produced by the feedback photodiode (TTI TIA-525I), with the input signal from the FPGA to drive the MZ modulator, with an additional amplification stage of $+27 \units{dB}$ (coaxial pulse amplifier ZPUL-30P) to span the entire $V_\pi$ interval of the modulator.

The SLED pump current is set to $250 \units{mA}$, in order to keep the optical power at the readout photodiode limited to $1 \units{mW}$ to ensure a linear response. The MZ modulator bias voltage is set to $1.6 \units{V}$, which yields a slightly shifted transfer function in order to compensate the input symbols distribution (see section \ref{subsubsec:cheqconst}). The optical attenuation can be set up to $100 \units{dB}$ with $0.01 \units{dB}$ precision. The attenuator is controlled by a Matlab script running on the computer.

\subsection{Input \& Readout}
\label{subsec:inout}

\begin{figure}
  \centering
  \includegraphics[width=0.45\textwidth]{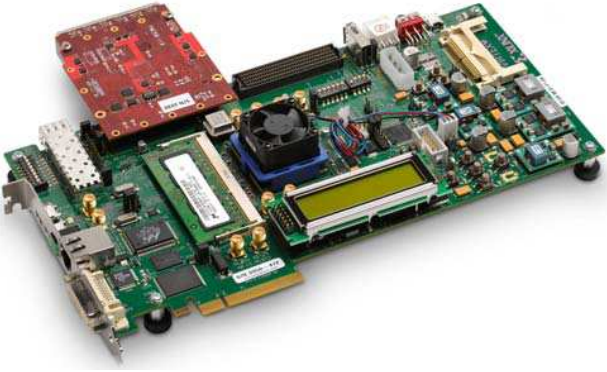}
  \caption{Xilinx ML605 board with Virtex 6 FPGA chip and 4DSP FMC150 daughter card (FMC150 and FMC151 cards look practically the same).}
  \label{fig:fpga}
\end{figure}

For our implementation, we use the Xilinx ML605 evaluation board (see figure \ref{fig:fpga}), powered by the Virtex 6 XC6VLX240T FPGA chip. The board is equipped with a JTAG port, used to load the FPGA design onto the chip, and a UART port, that we use to communicate with the board (as described in section \ref{sec:design}). The LPC (Low Pin Count) FMC (FPGA Mezzanine Card) connector is used to attach the 4DSP FMC151 daughter card, containing one two-channel ADC (Analog-to-Digital converter) and one two-channel DAC (Digital-to-Analog converter). The ADC's maximum sampling frequency is $250 \units{MHz}$ with 14-bit resolution, while the DAC can sample at up to $800 \units{MHz}$ with 16-bit precision. 

The synchronisation of the FPGA board with the reservoir delay loop is crucial for the performance of the experiment. For proper acquisition of reservoir states, the ADC has to output an integer number of samples per roundtrip time. The daughter card contains a flexible clock tree, that can drive the converters either from the internal clock source, or an external clock signal. As the former is limited to the fixed frequencies of the onboard oscillator, we employ the latter option. The clock signal is generated by a Hewlett Packard 8648A signal generator. With a reservoir of $N=51$ neurons (one neuron is added to desynchronise the inputs from the reservoir, as in \cite{paquot2012optoelectronic}) and a roundtrip time of $7.94 \units{\textmu s}$, the sampling frequency is set to $128.4635 \units{MHz}$, thus producing $20$ samples per reservoir state. To get rid of the transients, induced mainly by the finite bandwidths of the ADC and DAC, the 6 first and 6 last samples are discarded, and the neuron value is averaged over the remaining 8 samples.

The tensions of the electric signal to and from the mezzanine card need to be adjusted in order to achieve the most efficient interface without damaging the hardware. The DAC output voltage of $2 \units{V}_\text{p-p}$ is sufficient for this experiment, as typical voltages of the input signal range between $100 \units{mV}$ and $200 \units{mV}$. The ADC is also limited to $2 \units{V}_\text{p-p}$ input voltage. With settings described in the previous section, the output voltage of the readout photodiode doesn't exceed $1 \units{V}_\text{p-p}$.

\subsection{Experimental parameters}
\label{subsec:expparams}

To achieve the best performance, we scan the most influential parameters, which are: the input gain $\beta$, the decay rate $k$, the channel signal-to-noise ratio and the feedback attenuation, that corresponds to the feedback gain parameter $\alpha$ in equations \eqref{eq:rcevo2}. The first three parameters are set on the FPGA board, while the last one is tuned on the optical attenuator. The input gain $\beta$ is stored as a 18-bit precision real in $\left[0, 1\right[$ and was scanned in the $\left[0.1, 0.3\right]$ interval. The decay rate $k$ is an integer, typically scanned from $10$ up to $50$ in a few wide steps. The noise ratios were set to several pre-defined values, in order to compare our results with previous reports. The feedback attenuation was scanned finely between $4.5 \units{dB}$ and $6 \units{dB}$. Lower values would allow cavity oscillations to disturb the reservoir states, while higher values would not provide enough feedback to the reservoir. 
Table \ref{tab:gdparams} contains the values of parameters we used for the gradient descent algorithm (defined in section \ref{subsec:gd}).

\begin{table}
  \centering
  \caption{Gradient descent algorithm parameters}
  \begin{tabular}{|c|c|c|c|}
    \hline
    $\lambda_0$ & $\lambda_\text{min}$ & $\gamma$ & $k$ \\
    \hline
    $0.4$ & $0$ & $0.999$ & $10$ -- $50$\\
    \hline
  \end{tabular}
  \label{tab:gdparams}
\end{table}

\subsection{Experiment automation}

The experiment is fully automated and controlled by a Matlab script, running on a computer. It is designed to run the experiment multiple times over a set of predefined values of parameters of interest and select the combination that yields the best results. For statistical purposes, each set of parameters is tested several times with different random input masks, as defined in section \ref{subsec:rc}.

At launch, connections to the optical attenuator and the FPGA board are established, and the parameters on the devices are set to default values. After generating a set of random input masks, the experiment is run once and the elapsed time is measured. The duration of one run depends on the lengths of train and test sequences and varies from $6\units{s}$ to $12\units{s}$. This is considerably shorter than the offline-trained implementation \cite{paquot2012optoelectronic}, that required about $30\units{s}$. The script runs through all combinations of scanned parameters. For each combination, the values of the parameters are sent to the devices, the experiment is run several times with different input masks and the resulting error rates (see section \ref{sec:design}) are stored in the Matlab workspace. Once all the combinations are tested, the connections to the devices are closed and all collected data is saved to a file.

\section{FPGA design}
\label{sec:design}

The FPGA design is written in standard IEEE 1076-1993 VHDL language \cite{ieeevhdl, pedroni2004circuit} and compiled with Xilinx ISE Design Suite 14.7, provided with the board. We also used Xilinx ChipScope Pro Analyser to monitor signals on the board, mostly for debugging and testing.

\begin{figure}
  \centering
  \includegraphics[width=0.45\textwidth]{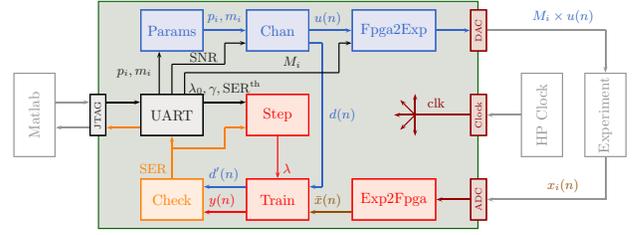}
  \caption{Simplified schematics of the FPGA design. The ML605 board is shown in green, the FMC151 card's components are rendered in maroon and other devices are coloured in grey. Smaller boxes and arrows inside the board represent modules (entities) and signals. The input layer modules (in blue) generate the target signal $d(n)$ and compute a nonlinear channel output $u(n)$. The readout layer (in red) receives the reservoir states $x_i(n)$ from the experiment, trains the weights $w_i$ and computes the output signal $y(n)$. The \pf{Check} module evaluates the symbol error rate. The \pf{UART} module executes commands issued by Matlab, sets variable parameters and sends the results back to the computer.}
  \label{fig:design}
\end{figure}

The simplified schematics of our design is depicted in figure \ref{fig:design}. Coloured boxes represent modules (i.e. entities) and the lines stand for data connections between them. As discussed in section \ref{subsec:inout}, the FPGA board implements both the input and the readout layers of the reservoir computer. Modules involved in each of these two functions are highlighted in blue and red, respectively. The board has a digital connection to a computer (running a Matlab script) and an analog one to the experimental setup. The former, realised through a UART port bridged to a standard COM port, is used to load parameters (e.g. $\lambda_0, \gamma, \ldots$) into the board and read the experiment results (i.e. symbol error rate) from the board. The latter consists of three analog connections: an output signal to the reservoir, containing the masked inputs $M_i \times u(n)$, a clock signal \pf{clk} from the HP signal generator and an input signal from the readout photodiode, containing reservoir states $x_i(n)$.

The operation of the FPGA board is controlled from the computer. A predefined set of 4-byte commands can be transmitted through the JTAG port, such as write a specific parameter value into the appropriate register or toggle the board state from reset to running, and vice versa. The commands are received and executed by the \pf{UART} module. In addition, when the FPGA is running, the module regularly transmits the value of the SER signal to the computer. In order to prevent collisions in the UART channel, commands from computer are only sent when the board is in a reset state, that is, no channel is being equalised.

The \pf{Chan} module implements the nonlinear channel model, given by equations \eqref{eq:qn} and \eqref{eq:chan}, and generates the input signal for the reservoir. It receives the noise amplitude, for a defined Signal-To-Noise ratio, from the computer via \pf{UART} module. The channel parameters $p_i$ and $m_i$ are supplied by the \pf{Params} module. Two Galois Linear Feedback Shift Registers (GLFSRs) with a total period of about $10^9$ are used to generate pseudorandom symbols $d(n) \in \left\{ -3,-1,1,3 \right\} $. Another GLFSR of period around $2\e{5}$ generates noise $\nu(n)$. The symbol sequence $d(n)$ is sent to the \pf{Train} module as a target signal, while the channel output $u(n)$ is multiplied by the input mask $M_i$ within the \pf{Fpga2Exp} module, and then converted to an analog signal by the FMC151 daughter card.

The analog reservoir output $x_i(n)$ is converted into a digital signal by the ADC. The time-multiplexed reservoir states are then sampled and averaged by the \pf{Exp2Fpga} module, which transmits all the neurons from one reservoir $\bar{x}(n)$ in parallel to the next module.

The synchronisation of the readout layer with the optoelectronic reservoir is performed by both \pf{Fpga2Exp} and \pf{Exp2Fpga} modules. At the beginning of a run of the experiment, the former sends a short pulse into the reservoir, before transmitting the input symbols. This pulse is detected by the \pf{Exp2Fpga} module and then used to synchronise the sampling and averaging process with the incoming reservoir states.

The \pf{Train} module implements the simple gradient descent algorithm. It receives the neurons $\bar{x}(n)$, the target signal $d(n)$ and the gradient step $\lambda$,  computes the reservoir output $y(n)$ with its error from the target signal, and adjusts the readout weights $w_i$ following equation \eqref{eq:wt}. The input target signal $d(n)$ is delayed by several periods $T$ to compensate the propagation time of the information through the input layer, the optoelectronic reservoir and the \pf{Exp2Fpga} module. The reservoir output $y(n)$ is then rounded up to the closest channel symbol $y(n) \leadsto \left\{-3,-1,1,3\right\}$ and compared to the delayed target signal $d'(n)$ by the \pf{Check} module, that counts misclassified symbols and outputs the resulting Symbol Error Rate.

The evolution of the learning rate $\lambda$ is governed by a separate module \pf{Step}, which implements the equation \eqref{eq:lambda_evo}, with initial value $\lambda_0$ and decay rate $\gamma$ set on the computer and transferred to the board through the UART connection. The module also monitors the performance of the reservoir computer and resets $\lambda$ to its initial value $\lambda_0$ when the Symbol Error Rate exceeds a predefined threshold value $\text{SER}^\text{th}$. This feature is used for the switching channel (see sections \ref{subsec:cheqswitch} and \ref{subsec:swch} ) and allows to improve the performance of the system by adjusting the readout weights to the new channel parameters.

The gradient descent algorithm is relatively simple, with only few addition and multiplication operations involved in equations \eqref{eq:wt} and \eqref{eq:lambda_evo}. While an adder can easily be built with a small amount of logic gates, multiplication is more complicated to implement and requires lots of resources. Moreover, as all readout weights are computed in parallel, the size of the design grows quickly with the number of neurons $N$. This results in slow implementation process and very low chances of generating a design that functions correctly. The solution resides in the use of special DSP48E slices, designed and optimised to perform a predefined set of arithmetic operations \cite{dsp48e1}. With proper settings, this dedicated microprocessor is capable of performing a $25\;\text{bit} \times 18\;\text{bit}$ multiplication in less than $6 \units{ns}$. While the speed gain compared to standard logic blocks is minimal, the implementation of the FPGA design is greatly simplified, as hundreds of logic gates and registers get replaced by just one component.

The arithmetic operations mentioned above are performed on real numbers. However, a FPGA is a logic device, designed to operate with bits. The performance of the design thus highly depends on the bit-representation of real numbers, i.e. the precision. The main limitation comes from the DSP48E slices, as these are designed to multiply a 25-bit integer by another 18-bit integer. To meet these requirements, our design uses a fixed-point representation with different bit array lengths for different variables. Parameters and signals that stay within the $\left]-1, 1 \right[$ interval are represented by 18-bit vectors, with 1 bit for the sign and 17 for the decimal part. These are the learning algorithm parameters $\lambda, \lambda_0$ and $\gamma$, the input mask elements $M_i$ and the reservoir states $x_i(n)$, extended from the 14-bit ADC output. Other variables, such as reservoir output $y(n)$ and readout weights $w_i$ span a wider $\left[-16,16\right]$ interval and are represented as 25-bit vectors, with 1 sign bit, 4 bits for the integer part and 20 bits for the decimal part.

Table \ref{tab:fpgares} reports total FPGA resource usage of our implementation. The design requires relatively few registers and Lookup Tables (LUTs). Most of the arithmetic operations are performed by the DSP48E slices, and their number grows roughly as $3 \times N$, thus theoretically limiting our reservoir to 255 neurons. Note that this restriction can be easily overcome by rearranging the DSP48E slices in a less concurrent design.  High internal memory (block RAM) usage is due to several ChipScope modules (not shown in figure \ref{fig:design}), added to monitor internal FPGA signals. To conclude, our implementation can be expanded to work with much bigger reservoirs.

\begin{table}
  \centering
  \caption{Total Usage of FPGA Resources}
  \begin{tabular}{|c|c|c|c|c|}
    \hline
     & Registers & LUTs & Block RAM & DSP48E \\
    \hline
    Used & 12288 & 5661 & 198 & 161 \\
    \hline
    Available & 301440 & 150720 & 416 & 768 \\
    \hline
    Utilisation & 4\% & 3\% & 47\% & 20\% \\
    \hline
  \end{tabular}
  \label{tab:fpgares}
\end{table}

\section{Results}
\label{sec:results}

This section presents the results of different investigations outlined in sections \ref{subsec:cheq} and \ref{subsec:gd}. All results presented here were obtained with the experimental setup described in section \ref{sec:expsetup}.

\subsection{Improved equalisation error rate}
\label{subsec:constch}

\begin{figure}
  \centering
  \includegraphics[width=0.44\textwidth]{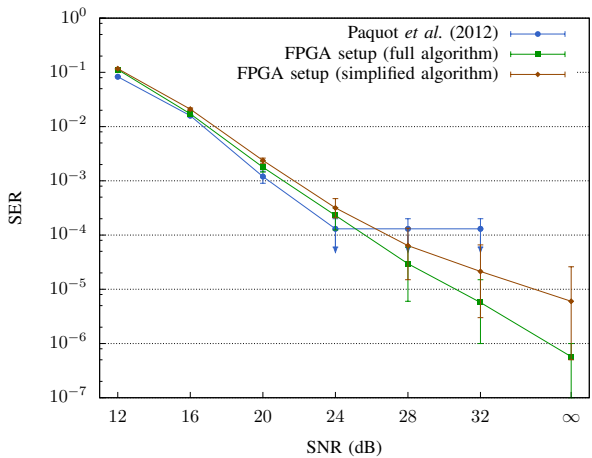}
  \caption{Experimental results obtained with our setup. Symbol Error Rates (SERs) are plotted against the Symbol-to-Noise Ratio (SNR). The equaliser was tested with 20 different random input masks over one million input symbols, average values are plotted on the graph (green squares). For the noiseless channel ($\text{SNR}=\infty$), for most choices of input mask, the RC made no errors over the test sequence. Blue dots show the results of the optoeletronic setup with offline training \cite{paquot2012optoelectronic}. For low noise levels, our system produces error rates significantly lower than \cite{paquot2012optoelectronic}, and for noisy channels the results are similar. Brown diamonds depict the SERs obtained with the simplified version of the training algorithm (see section \ref{subsubsec:simptrain}). The equalisation is less efficient than with the full algorithm, but the optimisation of experimental parameters takes less time.}
  \label{fig:oldvsfpga}
\end{figure}

Figure \ref{fig:oldvsfpga} presents the performance of our reservoir computer for different Signal-to-Noise Ratios (SNRs) of the wireless channel (green squares).
We investigated realistic SNR values for real world channels such as $60 \units{GHz}$ LAN \cite{wang2012capacity} and Wi-Fi \cite{duarte2014design}.
For each SNR, the experiment was repeated 20 times with different random input masks. Average SERs are plotted on the graph, with error bars corresponding to maximal and minimal values obtained with particular masks. We used noise ratios from $12\units{dB}$ up to $32\units{dB}$, and also tested the performance on a noiseless channel, that is, with infinite SNR. The RC performance was tested over one million symbols, and in the case of a noiseless channel the equaliser made zero error over the whole test sequence with most input masks.

\begin{figure}
  \centering
  \includegraphics[width=0.44\textwidth]{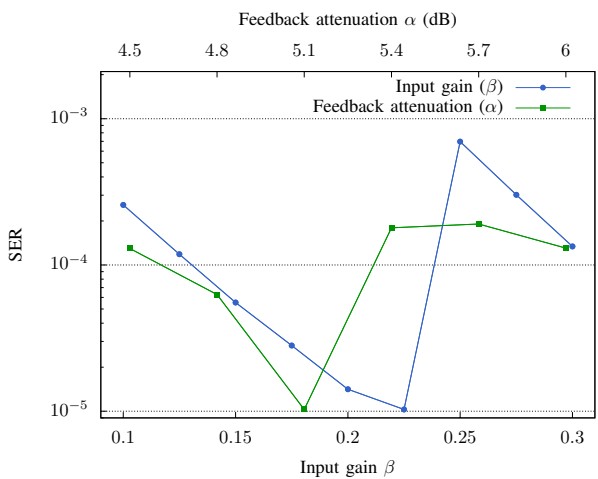}
  \caption{Dependence of the equaliser performance (at $32\units{dB}$ SNR) on the experimental parameters. Average SERs (over 10 random input masks) are plotted against the input gain (blue dots) and the feedback attenuation (green squares). The optimal feedback attenuation has to be set around $5.1 \pm 0.3 \units{dB}$, outside this region the SER deteriorates by roughly one order of magnitude. The input gain shows a minimum around $0.225 \pm 0.025$.}
  \label{fig:servsparams}
\end{figure}

The experimental parameters, such as the input gain $\beta$ and the feedback attenuation $\alpha$, were optimised independently for each input mask. Figure \ref{fig:servsparams} shows the dependence of the SER on these parameters. The plotted SER values are averaged over 10 random input masks. For this figure, we used data from a different experiment run with more scanned values. For each curve, the non-scanned parameter was set to the optimal value. The equaliser shows moderate dependence on both parameters, with an optimal input gain located within $0.225 \pm 0.025$ and an optimal feedback attenuation of $5.1 \pm 0.3 \units{dB}$.

We compare our results to those reported in \cite{paquot2012optoelectronic}, obtained with the same optoelectronic reservoir, trained offline (blue dots). For high noise levels ($\text{SNR} \leq 20\units{dB}$) our results are similar to those in \cite{paquot2012optoelectronic}. For low noise levels ($\text{SNR} \geq 24 \units{dB}$) the performance of our implementation is significantly better. Note that the previously reported results are only rough estimations of the equaliser's performance as the input sequence was limited by hardware to $6k$ symbols \cite{paquot2012optoelectronic}. In our experiment the SER is estimated more precisely over one million input symbols. For the lowest noise level ($\text{SER}=32\units{dB}$) an SER of $1.3\e{-4}$ was reported in \cite{paquot2012optoelectronic}, while we obtained an error rate of $5.71\e{-6}$ with our setup. 
One should remember that common error detection schemes, used in real-life applications, require the SER to be lower than $10^{-3}$ in order to be efficient.

To the best of our knowledge, the results presented here (at $32\units{dB}$ SNR) are the lowest error rates ever obtained with a physical reservoir computer. SERs around $10^{-4}$ have been reported in \cite{paquot2012optoelectronic, duport2012all, dejonckheere2014all} and a recently reported passive cavity based setup \cite{vinckier2015high} achieved a $1.66\e{-5}$ rate (this values is limited by the use of a $60k$-symbol test sequence), but no results below $10^{-5}$ have been published so far.
However, this isn't the main achievement of this experiment. Indeed, had it been possible to test \cite{paquot2012optoelectronic} on a longer sequence, it is possible that comparable SERs would have been obtained. The strength of this setup resides in the adaptability to changing environment, as will be shown in the following sections.

\subsection{Simplified training algorithm}
\label{subsec:ressimptrain}

The performance of the simplified training algorithm is shown in figure \ref{fig:oldvsfpga} (brown dots). The equaliser was tested with 10 random input masks and one million input symbols, the training was performed over $100k$ symbols. Only three parameters were scanned during these experiments: the input gain $\beta$, the feedback attenuation $\alpha$ and the signal-to-noise ratio. The learning rate $\lambda$ was set to $0.01$. The overall experimental runtime was significantly shorter: while an experiment with full training algorithm would last for about 50 hours, these results were obtain in approximately 10 hours (which is due to five different values of $k$ tested in the former case).

For high noise levels the results of the two algorithms are close and for low noise levels the simplified version yields slightly worse error rates. The performance is much worse in the noiseless case and strongly depends on the input mask: we notice a difference of almost two orders of magnitude between the best and the worst result. This performance loss is the price to pay for the simplified algorithm and shorter experimental runtime.

\subsection{Equalisation of a slowly drifting channel}
\label{subsec:driftchan}

Besides the environmental conditions, the relative positions of the emitter and the receiver can have a significant impact on the properties of a wireless channel. A simple example is a receiver moving away from the transmitter, causing the channel to drift more or less slowly, depending on the relative speed of the receiver. Here we show that our Reservoir Computer is capable of dealing with drifts with time scales of order of a second. This time scale is in fact slow compared to those expected in real life situations, but the setup could be sped up by several orders of magnitude, as will be shown in the next section.

A drifting channel is a good example of a situation where training the reservoir online yields better results than offline. We have previously shown in numerical simulations that training a reservoir computer offline on a non-stationary channel results in an error rate ten times worse than with online training \cite{antonik2015online}. We demonstrate here that an online-trained experimental reservoir computer performs well even on a drifting channel if $\lambda_{min}$ is set to a small non-zero value (see section \ref{subsubsec:gdnonstationary}).

At first, we investigated the relationship between the channel model coefficients and the lowest error rate achievable with our setup. That is, would the equalisation performance be better or worse if one of the numerical values in equations \eqref{eq:qn} and \eqref{eq:chan} was changed by, for instance, $10\%$. Given the vast amount of possibilities of varying the 4 parameters $p_i$ and $m$, we picked those that seemed most interesting and most significant. We thus tested the amplitude of the linear part, given by the parameter $p_1$, the amplitude of the quadratic and cubic parts, given by $p_2$ and $p_3$, and the memory $m$ of the impulse response. For each test, only one aspect of the channel was varied and other parameters were set to default values (as in equations \eqref{eq:qn} and \eqref{eq:chan}). The results of these investigations are presented in the Appendix.

We then programmed these parameters to vary during experiments in two different ways: a monotonic growth (or decay) and a periodic linear oscillation between two defined values. The results of these experiments are depicted in figure \ref{fig:driftchan}.

Figure \ref{fig:driftchan}(a) shows the experimental results for the case of monotonically decreasing $p_1$ from $1$ to $0.652$. The blue curve presents the resulting SER with $\lambda_{min}=0$, that is, with training process stopped after $45k$ input symbols. The green curve depicts the error rate obtained with $\lambda_{min} = 0.01$, so that the readout weight can be gradually adjusted as the channel drifts. Note that while in the first experiment the SER grows up to $0.329$, it remains much lower in the second case. The increasing error rate in the latter case is due to the decrease of $p_1$ resulting in a more complex channel. Brown curves show the best possible error rate obtained with our setup for different values of $p_1$, as presented in the Appendix. With $p_1$ approaching $0.652$, the obtained error rate is $8.0\e{-3}$, which is the lowest error rate possible for this value of $p_1$, as demonstrated in figure \ref{subfig:p1ser}. This shows that the non-stationary version of the training algorithm allows a drifting channel to be equalised with the lowest error rate possible.

\newlength{\octafigheight}
\setlength{\octafigheight}{4.5cm}

\begin{figure*}
  \centering
  \begin{tabular}{ll}
    \includegraphics[height=\octafigheight]{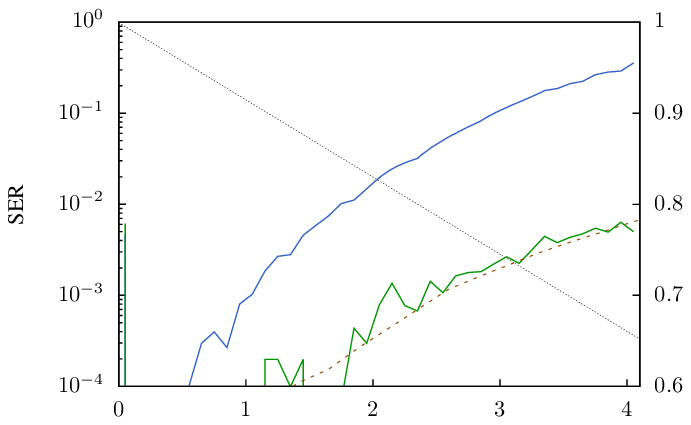}
    \begin{picture}(0,0) \put(-225,115){(a)} \end{picture} &
    \includegraphics[height=\octafigheight]{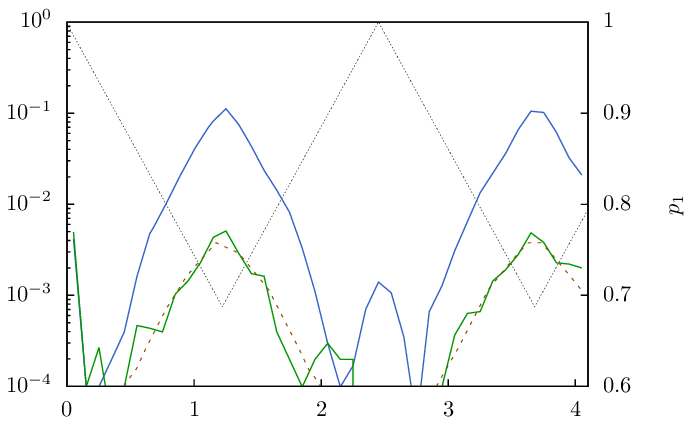}
    \begin{picture}(0,0) \put(-230,115){(b)} \end{picture} \\
    \includegraphics[height=\octafigheight]{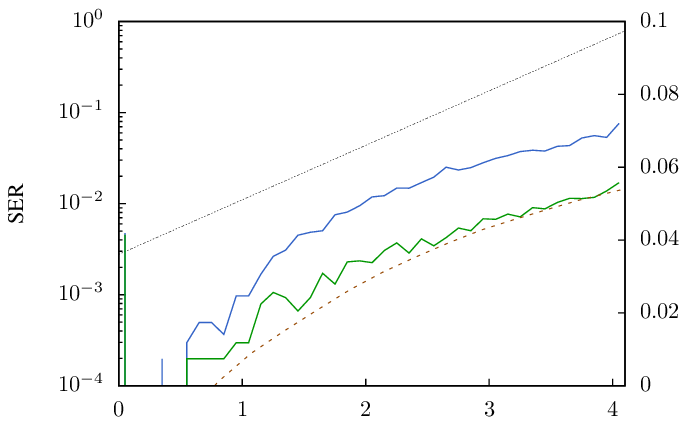}
    \begin{picture}(0,0) \put(-225,115){(c)} \end{picture} &
    \includegraphics[height=\octafigheight]{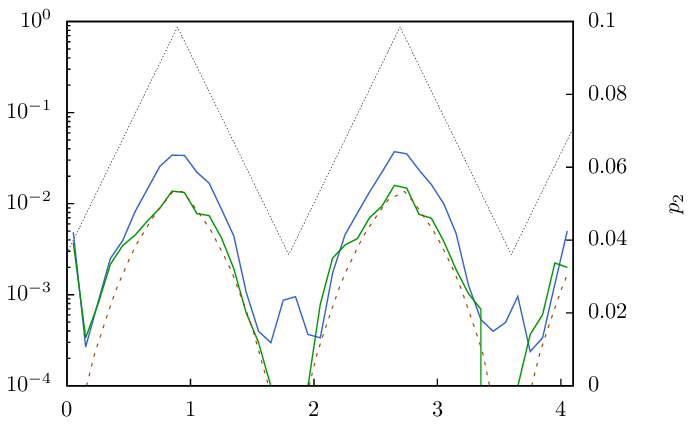}
    \begin{picture}(0,0) \put(-230,115){(d)} \end{picture} \\
    \includegraphics[height=\octafigheight]{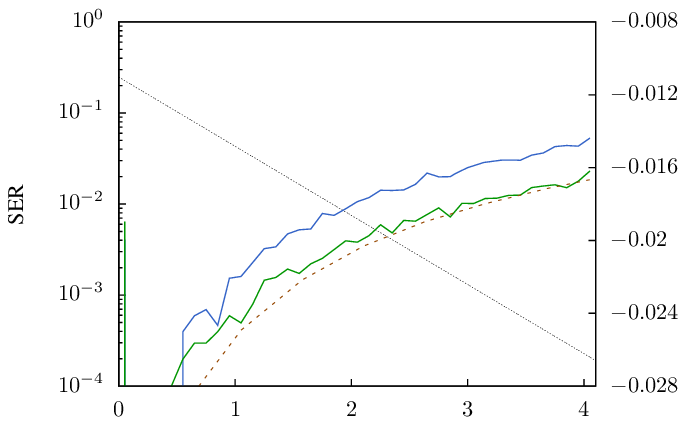}
    \begin{picture}(0,0) \put(-225,115){(e)} \end{picture} &
    \includegraphics[height=\octafigheight]{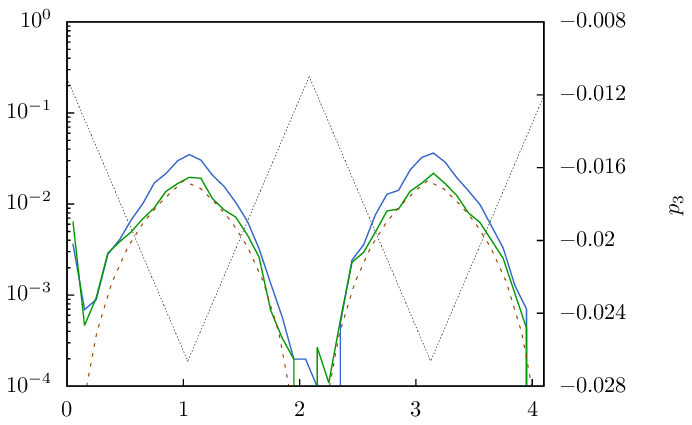}
    \begin{picture}(0,0) \put(-230,115){(f)} \end{picture} \\
    \includegraphics[height=\octafigheight]{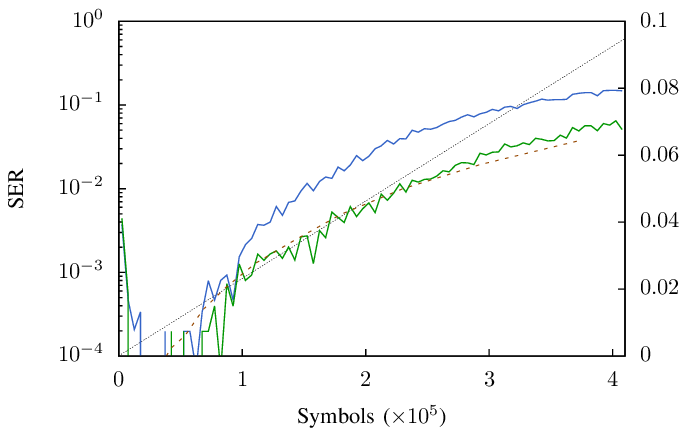}
    \begin{picture}(0,0) \put(-225,115){(g)} \end{picture} &
    \includegraphics[height=\octafigheight]{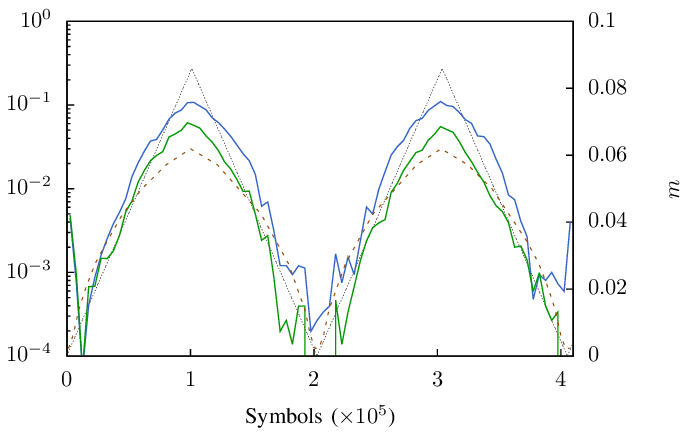}
    \begin{picture}(0,0) \put(-230,115){(h)} \end{picture} \\
  \end{tabular}
  \caption{Symbol error rates (right axis, log scale), averaged over $10k$ symbols, produced by the experimental setup with a drifting channel. Each panel presents data obtained from one experiment run with a fixed input mask and optimal parameters $\alpha$, $\beta$ and $k$. Two different training methods were tested: blue curves show the results produced by the full training algorithm with $\lambda_{min}=0$ (see section \ref{subsubsec:gdfull}), while green curves depict those obtained with the non-stationary version with $\lambda_{min} > 0$ (see section \ref{subsubsec:gdnonstationary}). Dashed brown lines display the best performance that can be obtained with our system for given values of variable parameters $p_i$ and $m$ (right axis, linear scale), shown in black (see Appendix for details).
  \textbf{(a) \& (b)} Monotonically decreasing and oscillating $p_1$. 
  \textbf{(c) \& (d)} Monotonically increasing and oscillating $p_2$. 
  \textbf{(e) \& (f)} Monotonically decreasing and oscillating $p_3$. 
  \textbf{(g) \& (h)} Monotonically increasing and oscillating $m$.}
  \label{fig:driftchan}
\end{figure*}

Figure \ref{fig:driftchan}(b) depicts error rates obtained with $p_1$ linearly oscillating between $1$ and $0.688$.
With $\lambda_{min}=0$ (blue curve) the error rate is as low as $1\e{-4}$ when $p_1$ is around $1$, and grows very high elsewhere. With $\lambda_{min}=0.01$, the obtained SER is always at the lowest value possible: at the point where $p_1=0.688$, it stays at $5.0\e{-3}$, which again is close to the best performance for such channel, illustrated by the brown curve.

We obtained similar results with parameters $p_2$, $p_3$ and $m$, as shown in figures \ref{fig:driftchan}(c)-(d). Letting the reservoir computer adapt the readout weights by setting $\lambda_{min}>0$ produces the lowest error rates possible for a given channel, while stopping the training with $ \lambda_{min}=0$ results in quickly growing SERs.

\subsection{Equalisation of a switching channel}
\label{subsec:swch}

\begin{figure}[t]
  \begin{center}
    \includegraphics[width=0.44\textwidth]{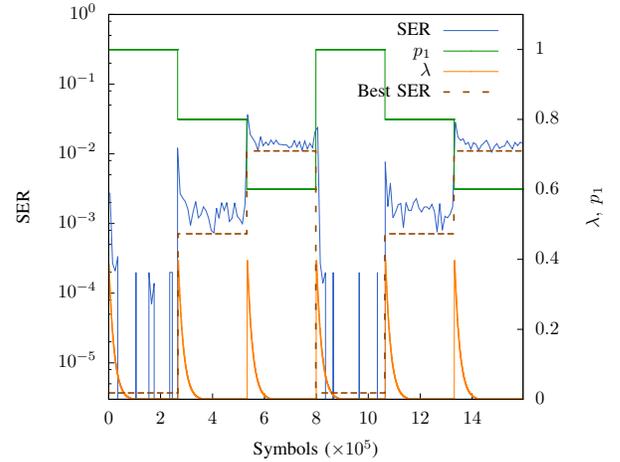}
  \end{center}
  \caption{Symbol error rate (left axis), averaged over $10k$ symbols, produced by the FPGA in case of a switching channel. The value of $p_1$ (right axis, green curve) is modified every $266k$ symbols. The change in channel is followed immediately by a steep increase of the SER. The $\lambda$ parameter (right axis, orange curve) is automatically reset to $\lambda_0 = 0.4$ every time a performance degradation is detected, and then returns to its minimum value, as the equaliser adjusts to the new channel, bringing down the SER to its asymptotic value. After each variation of $p_1$, the reservoir re-trains. The lowest error rate possible for the given channel is shown by the dashed brown curve.}
  \label{fig:swch}
\end{figure}

Figure \ref{fig:swch} shows the error rate produced by our experiment in case of a switching noiseless communication channel. The parameters of the channel are programmed to switch in cycle among equations \eqref{eq:varch} every $266k$ symbols. Every switch is followed by a steep increase of the SER, as the reservoir computer is no longer optimised for the channel it is equalising. The performance degradation is detected by the algorithm, causing the learning rate $\lambda$ to be reset to the initial value $\lambda_0$, and the readout weights are re-trained to new optimal values.

For each value of $p_1$, the reservoir computer is trained over $45k$ symbols, then its performance is evaluated over the remaining $221k$ symbols. In case of $p_1=1$, the average SER is $1\e{-5}$, which is the expected result. For $p_1=0.8$ and $p_1=0.6$ we compute average SERs of $7.1\e{-4}$ and $1.3\e{-2}$, respectively, which are the best results achievable with such values of $p_1$ according to our previous investigations (see figure \ref{subfig:p1ser}). This shows that after each switch the readout weights are updated to new optimal values, producing the best error rate for the given channel.

Note that the current setup is rather slow for practical applications. With a roundtrip time of $T=7.94 \units{\textmu s}$, its bandwidth is limited to $126 \units{kHz}$ and training the reservoir over $45k$ samples requires $0.36 \units{s}$ to complete. However, it demonstrates the potential of such systems in equalisation of non-stationary channels. For real-life applications, such as for instance Wi-Fi 802.11g, a bandwidth of $20 \units{MHz}$ would be required. This could be realised with a $15 \units{m}$ fibre loop, thus resulting in a delay of $T=50 \units{ns}$. This would also decrease the training time down to $2.2 \units{ms}$ and make the equaliser more suitable for realistic channel drifts. 
The speed limit of our setup is set by the bandwidth of the different components, and in particular of the ADC and DAC. For instance with $T=50\units{ns}$ and keeping $N=50$, reservoir states should have a duration of $1\units{ns}$, and hence the ADC and DAC should have bandwidths significantly above $1\units{GHz}$ (such performance is readily available commercially). As an illustration of how a fast system would operate, we refer to the optical experiment \cite{brunner2012parallel} in which information was injected into a reservoir at rates beyond $1\units{GHz}$.

\section{Conclusion}
\label{sec:ccl}

In the present work we applied the online learning approach to training an opto-electronic reservoir computer. We programmed the simple gradient descent algorithm on an FPGA chip and tested our system on the nonlinear channel equalisation task. We obtained error rates up to two orders of magnitude lower than previously reported RC implementations on the channel equalisation task, while significantly reducing the experimental runtime. 

We also demonstrated that our system is well-suited for non-stationary tasks by equalising a drifting and a switching channel. In both cases, we obtained the lowest error rates possible with our setup. Such flexibility is more complex to achieve with offline methods, and would require improving the algorithm by adding several computational steps. The online learning methods, on the other hand, need little modifications to successfully solve this task. Moreover, in case of a slowly drifting channel the algorithm can be set to fine-tune the readout weights without performing a complete re-training of the reservoir, which would be hard to achieve with offline learning. This shows that the technique presented here is more suitable for real-life tasks with variable parameters.

Our realisation opens several new research directions. Using the FPGA to drive the opto-electronic reservoir gives more control over the experiment. Such a system could, for instance, implement a full optimisation of the readout weights and the input mask, as suggested in \cite{hermans2015optoelectronic, hermans2015photonic}. The real-time training makes it possible to feed the output signal back into the reservoir.
This additional feedback would highly enrich the dynamics of the system, allowing one to tackle new tasks such as pattern generation or chaotic series prediction \cite{antonik2016towards}.
The high speed of dedicated electronics offers the opportunity to develop very fast, autonomous reservoir computers with GHz data rates. The present work thus paves the way towards autonomous, very-high speed, fully analog reservoir computers with a wider range of possible applications.

\appendix[Influence of channel model parameters on equaliser performance]
\label{sec:inflparams}

\begin{figure*}
  \centering
  \subfigure[]{\includegraphics[height=\octafigheight]{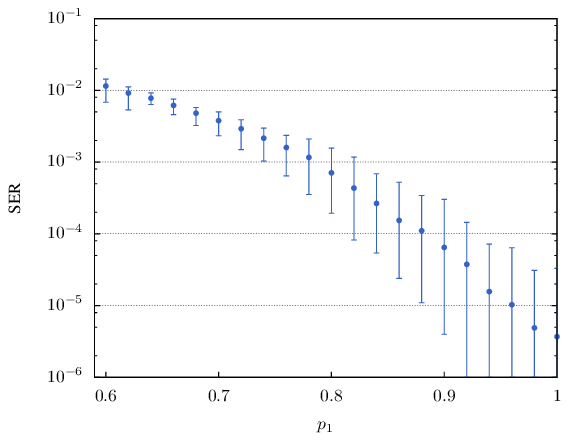}\label{subfig:p1ser}}
  \subfigure[]{\includegraphics[height=\octafigheight]{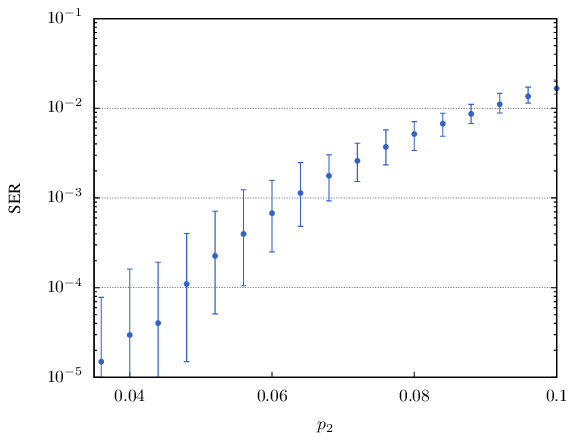}\label{subfig:p2ser}}
  \subfigure[]{\includegraphics[height=\octafigheight]{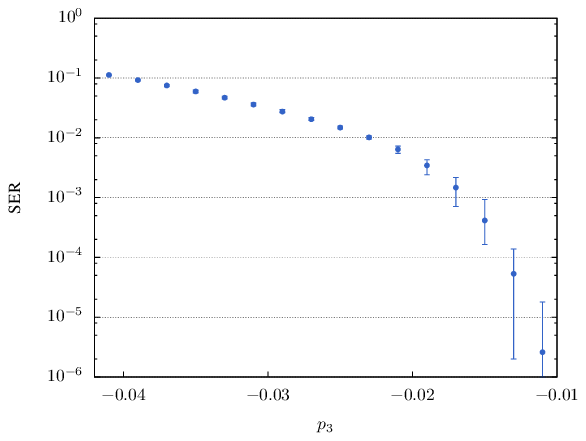}\label{subfig:p3ser}}
  \subfigure[]{\includegraphics[height=\octafigheight]{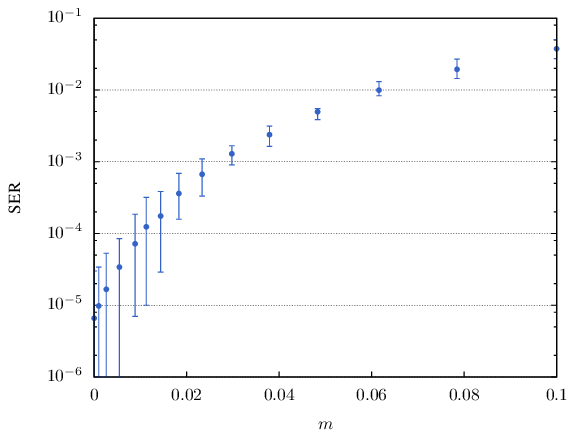}\label{subfig:mmser}}
  \caption{Error rates for different values of channel parameters $p_i$ and $m$. The results were measured over one million input symbols, with 10 random input masks and zero noise. \textbf{(a)} Lower $p_1$ implies lower linear part of the channel, containing the useful signal, which naturally results in higher error rates. \textbf{(b)} Increasing the quadratic component $p_2$ of the channel makes it more nonlinear, and thus more difficult to equalise. \textbf{(c)} Increasing the cubic component $|p_3|$ of the channel makes it more nonlinear, and thus more difficult to equalise. \textbf{(d)} Higher values of $m$ make the channel equalisation more complex.}
\end{figure*}

Figure \ref{subfig:p1ser} shows the equalisation results for different values of $p_1$. We tested each value over 10 random input masks, with independent experimental parameters optimisation for each run. Average values are presented on the plot, with error bars depicting best and worst results obtained among different masks. The equaliser performance was tested on a sequence of one million inputs, and in several cases we obtained zero misclassified symbols. Note that the observed increase of the SER with reduction of $p_1$ is natural as the linear part contains the signal to be extracted. When decreasing $p_1$, not only the useful signal gets weaker, but the nonlinear distortion also becomes relatively more important.

Figures \ref{subfig:p2ser} and \ref{subfig:p3ser} present the dependence of the SER on parameters $p_2$ and $p_3$, respectively. These parameters define the amplitude of the nonlinear distortion of the signal, and as they grow, the channel becomes more nonlinear and thus more difficult to equalise. The results of equalisations with different values of $m$ are shown in figure \ref{subfig:mmser}, higher values of $m$ increase the temporal symbol mixing of the channel, hence worse results.

\section*{Acknowledgment}

We acknowledge financial support by Interuniversity Attraction Poles program of the Belgian Science Policy Office under grant IAP P7-35 “photonics@be”,  by the Fonds de la Recherche Scientifique FRS-FNRS and by the Action de la Recherche Concert\'{e}e of the Acad\'{e}mie Universitaire Wallonie-Bruxelles under grant AUWB-2012-12/17-ULB9.

\ifCLASSOPTIONcaptionsoff
  \newpage
\fi



\end{document}